\documentclass[11pt,a4paper]{article}
\usepackage[T1]{fontenc}
\usepackage{tgbonum}
\usepackage{amsmath,slashed,graphicx,booktabs,array,url,hyperref}
\usepackage[leftcaption]{sidecap}
\sidecaptionvpos{figure}{c}
\newcommand{\ft}[2]{{\textstyle\frac{#1}{#2}}}
\def\di{\slashed{D}}
\def \ha{\frac{1}{2}}
\def \pa {\partial}
\def \eps {\epsilon}
\def \trc {{\mathcal T}r\chi}
\def \S {{\mathcal S}}
\def \C {{\mathcal C}}

\def \R {{\mathcal R}}

\def \H {{\mathcal H}}

\def \al {\alpha}

\def \aip {\partial^{\al }_{\xi } a_k }

\def \qp {D^{\al }_x q_{-m-l} }
\def \ddp {D^{\al }_x d_{-2-j} }
\def \qp {D^{\al }_x q_{-2-j }}
\def \lap {\big(-\varDelta+\al{\mathcal R}+{i\lambda \over t }\big)}
\def \ti {\tilde }
\def \d { d_{-5}}
\def \tri { \varDelta  \varphi }
\def \trii { \varDelta ^{(3)} \varphi }
\def \RM { \partial {\mathcal M } }
\def \M {{\mathcal M }}
\def \kab {{\mathcal K} _{ab}}
\def \kbb {{\mathcal K} _{bb}}
\def \trk {{\mathcal T}r{\mathcal K}}
\def \trkk {{\mathcal T}r{\mathcal K}^2 }
\def \trkkk {{\mathcal T}r {\mathcal K}^3 }
\def \chr {\Gamma ^c_{ab}}

\def \k {{\mathcal K}}
\def \prr {\partial^2_r }
\def \prrr {\partial^3_r }
\def \pr {\partial_r }
\def \tchr {\tilde\Gamma ^c_{ab}}
\def \tchrr {\tilde\Gamma ^c_{ab,r}}
\def \tchrrr {\tilde\Gamma ^c_{ab,rr}}
\def \gab { g _{ab}}
\def \goab {g^{ab}}
\def \gabr {g_{ab,r}}

\def \ggr {g^{ab}g_{ab,r}}
\def \ggoaarbb { g^{aa}_{,rbb}}
\def \gabpp {g^{ab} \partial _a \partial _b }
\def \gchrx {g^{ab} \Gamma ^c _{ab} \xi _c }
\def \ggoaabb { g^{aa}_{,bb}}
\def \ggaarr { g_{aa,rr}}
\def \ggoaarr { g^{aa}_{,rr}}
\def \ggaarrr { g_{aa,rrr}}
\def \ggoaarrr { g^{aa}_{,rrr}}

\def \ggoabab { g^{ab}_{,ab}}
\def \ggoabrab { g^{ab}_{,rab}}
\def \tgoab {\tilde g^{ab}}
\def \tgoabr {\tilde g^{ab}_{,r}}
\def \tgabrr {\tilde g_{ab,rr}}
\def \tgabrrr {\tilde g_{ab,rrr}}
\def \tgoabrr {\tilde g^{ab}_{,rr}}

\def \pxi{\partial _{\xi}}
\def \pox{\partial ^x}
\def \poxi{\partial ^{\xi}}
\def \t {\tau }
\def \si{\sigma }
\def \lam{\lambda }
\def \ro{\varrho}
\def \xia{\xi _a }
\def \xib{\xi _b }
\def \xic{\xi _c }
\def \ex{e^{-\ti \rho r }}
\begin{document}
	\begin{titlepage}
		\title {
			\hfill{\small ETH-TH/91-19, MPI-PAE-PTH/45-91}\\[5mm]
			Finite Size Effects from General
			Covariance and Weyl Anomaly}
		\author{A. Dettki\\[1mm]
			\emph{\small Max-Planck-Institut f\"ur Physik und Astrophysik}\\
			\emph{\small(Werner-Heisenberg-Institut f\"ur Physik)}\\
			\emph{\small P.O.Box 40 12 12, Munich (Germany)}
			\and 
		     A. Wipf\footnote{present address: 
		     	Theoretisch-Physikalisches-Institut, Friedrich-Schiller-Universit\"at Jena, 07743 Jena,
		     	Germany; email:
		     	wipf@tpi.uni-jena.de; url:
		     	\url{http://www.tpi.uni-jena.de/qfphysics/homepage/wipf/index.html}}\\[1mm]
			\emph{\small Institut f\"ur Theoretische Physik,
		Eidgen\"ossische Technische Hochschule}\\
		\emph{\small H\"onggerberg, Z\"urich CH-8093, Switzerland}}
		\date{\small 12 August 1991; Latex-ed: 12 January 2022}
		\maketitle
\begin{abstract}
			\noindent
By exploiting the diffeomorphism invariance we relate the
finite size effects of massless theories to their Weyl anomaly.
We show that the universal contributions to the finite size
effects are determined by certain coefficient functions in the heat
kernel expansion of the related wave operators. For massless
scalars confined in a $4$-dimensional curved spacetime with
boundary the relevant coefficients are given -- confirming the results
of Moss and Dowker and also of Branson and Gilkey.
We apply the general results to theories on bounded regions in two- and
four-dimensional flat space-times and determine the change of the
effective action under arbitrary conformal deformations of the regions.
\end{abstract}
\vskip2mm
\begin{center}\small
 \\[5mm]
 Published version: Nuclear Physics \textbf{B377} (1992) 252-280\\
 doi: 10.1016/0550-3213(92)90024-6\\
 arXiv-ed August 10, 1991 ; \LaTeX-ed January 12, 2022
\end{center}
\tableofcontents
\end{titlepage}

\section{Introduction}\label{sec:introduction}
Finite size effects play an important role for critical
systems having no intrinsic length scale except those
dictated by the geometry. They are caused by the dependence
of the vacuum or equilibrium state on the underlying space-time.
For example, the energy-momentum of the vacuum depends
on the boundary enclosing the systems and this leads
to a measured Casimir force acting on the boundary.
The geometry-dependence appears as
anomalous contributions to the effective action
which generates the correlation functions of the energy-momentum tensor.
The anomalies are due to external gravitational
fields and/or boundaries of space-time.
Their consequences have been investigated in a wide range of theories
like QED \cite{1}, QCD \cite{2}, Kaluza-Klein theories \cite{3},
2-dimensional conformal field theories \cite{4} and Stringtheories \cite{5}.

Actually the gravitational- and boundary anomalies are related
by general covariance and this interrelation will be considerably
exploited in our investigations of finite size effects.
In what follows general covariance plays an essential role
and thus we choose a manifestly covariant regularization, namely
the zeta-function regularization which immediately connects to Schwingers
proper time (heat kernel) expansion. This regularization scheme
is convenient to extract the geometry dependence of expectation
values and in particular the relation between the bulk- and surface
terms. One only needs to separate the bulk- and surface contributions
to the heat kernel expansion for the wave operators of interest.
Recently much efforts have been made to work out the relevant
expansion coefficients for various field theories
and different boundary conditions \cite{7,8,9,10}. Our surface terms
agree with the earlier results of Dowker and Moss \cite{8} and Branson
and Gilkey \cite{9}.\par
In this paper we study massless particles which are Weyl-invariantly
coupled to gravity and which are confined in a finite
space-time region. Although we are mainly concerned with finite size
effects in flat Euklidean space-time it pays off to
couple the particles to a gravitational field and only assume
space-time to be euklidean at the end of the computations. This allows
us to exploit the consequences of general covariance which relates
the bulk- and surface terms. For example, the identification
of the central charge of two-dimensional models (defined via the
short distance behaviour of the energy-momentum correlators which
is determined by the bulk term of the effective action) as
Casimir effect (determined by the surface term of the effective action)
follows immediately when one couples the system to gravity.\par
Our results apply to arbitrary massless particles interacting
with the gravitational field and the boundary. Since for
different particles the finite size effects
are related if certain constants in the heat kernel expansion are adjusted
accordingly we give the explicit results for scalar particles only,
since they play a prominent role in the inflationary cosmological scenarios.
\par
The paper is organized as follows. In section \ref{sec:finite1}
the significance of the Weyl anomaly for finite size effects
and the interplay between bulk and surface terms for
conformal field theories is discussed. In particular we
relate certain coefficients in the heat kernel expansion
to the finite size effects. In the following section 
\ref{sec:finite2}
these general results are applied to $2$-dimensional systems
and the response of the quantum system to arbitrary
changes of the boundary is derived. In section \ref{sec:heat}
we outline the computation of the relevant heat kernel coefficients
for $4$-dimensional curved space-times with boundaries. We used Seeley's
method \cite{6} to determine these Seeley-deWitt coefficients. This project
has been undertaken independently from \cite{7,8,9}, has not yet
been published and
was only privately communicated \cite{10}. But the trilogy of papers
\cite{7,8,9} makes clear, that it is worth having several
derivations of this important result obtained by different methods.
In the last section \ref{sec:applications} the general results are applied to $4$-dimensional
systems and the finite size effects for simple geometries are evaluated.
In appendix $A$ the notation and conventions used in the main body
of the paper are explained and in appendix $B$ all relevant
Seeley-deWitt coefficients in $4$ dimensions for scalar particles
obeying Dirichlet boundary conditions are listed. The reader who
is less interested in technical details may skip part of 
section \ref{sec:heat} and take formula \eqref{4.52} as main result of this section.

\section{Finite Size Effects From Weyl Anomaly}\label{sec:finite1}
In this paper we shall investigate the change of
field theoretical quantities under conformal transformations.
A {\sl conformal transformation} $f:\{{\M},g\}\to\{{\mathcal N},\tilde g\}$
is a map that preserves angles but not necessarily
distances. The spacetimes ${\mathcal M}$ and ${\mathcal N}$ may possess
boundaries $\pa{\mathcal M}$ and $\pa {\mathcal N}$ and for
simplicity we shall assume that both are submanifolds of the same
$d$-dimensional Lorentzian (Riemannian) spacetime $X$
and their boundaries are hypersurfaces in $X$. Then $g$ and $\tilde g$
are the metric of $X$ restricted to ${\mathcal M}$ and ${\mathcal N}$, respectively,
and in the following both are denoted by $g$.
Since $f$ leaves angles invariant the distance between neighbouring
points can only change by a local scale factor. Choosing
local coordinates on $X$, so that
\begin{equation}
f:\{{\M},\pa{\M}\}\longrightarrow \{{\mathcal N},\pa {\mathcal N}\};
\quad x^\mu\longrightarrow y^\mu=f^\mu(x)\label{2.1}
\end{equation}
this means that
\begin{equation}
g_{\mu\nu}(y)\,dy^\mu dy^\nu=e^{2\varphi(x)}
g_{\mu\nu}(x) dx^\mu dx^\nu,\label{2.2a}
\end{equation}
where the local scale factor is determined by the metric
and conformal transformation as
\begin{equation}
e^{2\varphi(x)}={1\over d}\,g_{\mu\nu}\big(y(x)\big){\pa y^\mu\over
\pa x^\sigma}{\pa y^\nu\over \pa x^\rho}\,g^{\sigma\rho}(x).\label{2.2b}
\end{equation}
It is important to distinguish between conformal
transformations and {\sl diffeomorphism}. A map $f:{\mathcal M}\to{\mathcal N}$
is conformal with respect to prescribed geometries
on ${\mathcal M}$ and ${\mathcal N}$ and the proper length may change
by a (local) scale factor. Of course, we may also interpret such an $f$ as
diffeomorphism (or coordinate transformation), but then the metric is
carried along, i.e.
\begin{equation}
\hat g_{\mu\nu}dy^\mu dy^\nu=g_{\mu\nu}dx^\mu dx^\nu \Longrightarrow
\hat g_{\mu\nu}(y)=e^{-2\varphi}g_{\mu\nu}(y)\label{2.2c}
\end{equation}
and differs from the prescribed metric on ${\mathcal N}$ by a Weyl
factor. Thus, an arbitrary conformal transformation $f$ is a
composition of a diffeomorphism (defined by the same $f$) and
a compensating Weyl transformation. Note in particular that the conformal
group is not a subgroup of the diffeomorphism group
since by diffeomophisms we mean maps \eqref{2.1} together with the
associated transformations of the metric tensor, $g=f^*\hat g$,
and matter fields. Also note that contrary to the
diffeomorphism group which is always infinite dimensional, it may
happen that there are no conformal maps from ${\mathcal M}$ to ${\mathcal N}$.\par
In Minkowski space-time the conformal transformations consist of
translations, Lorentz transformations, dilatations
\begin{equation}
y^\mu=\lambda x^\mu\label{2.3a}
\end{equation}
and special conformal transformations
\begin{equation}
y^\mu={x^\mu+x^2 b^\mu\over 1+2b\cdot x+b^2 x^2}\label{2.3b}
\end{equation}
and form a $SO(d,2)$ (in Euklidean space a $SO(d+1,1)$) group. The
scale factor is one for the Poincare subgroup and it is
\begin{equation}
e^\varphi=\lambda,\qquad e^\varphi=\big(1+2b\cdot x+b^2x^2\big)^{-1}\label{2.3c}
\end{equation}
for the dilatations and special conformal transformations, respectively.\par
Let us now assume that a massless matter field $\Phi_{\mu\nu\dots}$ (which
may be a spinor or tensor field) couples Weyl invariantly to the gravitational field,
i.e. that the classical action is invariant under the Weyl transformations
of the metric and matter fields
\begin{equation}
\big\{x^\mu,g_{\mu\nu}(x),\Phi_{\mu\nu\dots}(x)\big\}\longrightarrow
\big\{x^\mu,e^{2\varphi(x)}g_{\mu\nu}(x),e^{\al\varphi(x)}\Phi_{\mu\nu\dots}(x)\big\}.
\label{2.4}
\end{equation}
For example, a massless scalar has Weyl weight $\al=\ha (2-d)$, a spinor $\al=
\ha (1-d)$ and a photon in $d=4$ dimensions has $\al=0$. For fermions
one needs to introduce a $d$-bein field. It is understood
that the $d$-bein inherits its Weyl transformation from the metric
field, i.e. $e^a_{\;\mu}\to e^\varphi e^a_{\;\mu}$.\par
Since a conformal transformation is a composition of a diffeomorphism
and a compensating Weyl transformation a generally covariant and Weyl
invariant classical field theory is automatically conformally invariant.
We prefer to change the order and first Weyl transform and then act with
the diffeomorphism $f$ in order for the Weyl transformation to act on
the original manifold $\M$ rather then on ${\mathcal N}$.
Thus, given a conformal map \eqref{2.1} and the correponding
Weyl factor \eqref{2.2b}, we first Weyl transform the metric
and matter fields with this Weyl factor and then act with
$f$, now interpreted as diffeomorphism. The net result
is the conformal transformation
\begin{equation}
\big\{x^\mu,g_{\mu\nu}(x),\Phi_{\mu\nu\dots}(x)\big\}
\longrightarrow \big\{y^\mu,g_{\mu\nu}(y),e^{\al\varphi}
\Lambda_\mu^{\;\sigma}\Lambda_\nu^{\;\rho}\dots
\Phi_{\sigma\rho\dots}(y)\big\}\,,\label{2.5}
\end{equation}
Since Weyl transformations and diffeomorphisms are
both classical symmetries this shows that conformal maps \eqref{2.5} are
indeed classical symmetries.
For example, in Minkowski space-time \eqref{2.5} leaves $\eta_{\mu\nu}$
invariant and are the well-known conformal symmetry transformations of
a Minkowskian field theory.
\par
Note that under an infinitesimal conformal transformation
\begin{equation}
y^\mu=x^\mu-\eps X^\mu(x),\qquad
X_{\mu;\nu }+ X_{\nu;\mu}={2\over d}\, g_{\mu \nu}
\nabla\cdot X
\label{2.6a}
\end{equation}
($X$ is a conformal Killing field for conformal transformations)
the matter field transforms as
\begin{equation}
\Phi_{\mu\nu\dots}\longrightarrow \big(L_X-{\al\over d}\nabla\cdot X^\big)
\Phi_{\mu\nu\dots}.\label{2.6b}
\end{equation}

Which of these classical symmetries survive in the quantum theory
depends on the chosen regularization. We shall use the manifestly
covariant zeta-function regularization such that the theory
is diffeomorpism invariant.
However, it is well-known that classical Weyl invariance
ceases to be a symmetry of a covariantly quantized theory if space time
is curved and/or has boundaries. This implies then that
the conformal invariance is broken as well.
In particular the change of the effective quantum action under conformal
transformations \eqref{2.1} is equals to the change under the corresponding
Weyl transformation \eqref{2.4}
\begin{equation}
\delta \Gamma\equiv\Gamma[{\mathcal N},g]-\Gamma[\M,g]=
\Gamma[\M,e^{2\varphi}g]-\Gamma[\M,g]
\label{2.7}
\end{equation}
if $\varphi$ is related to the conformal transformation \eqref{2.1} by
\eqref{2.2b}.\par
The variation of $\Gamma$ under Weyl transformations is determined
by the Weyl anomaly (or trace anomaly of the
energy-momentum tensor). This anomaly is local in the curvature
of space-time and its covariant derivatives and in the
extrinsic and intrinsic curvature of the boundary. Is is determined
by the $t$-independent term in the expansion of the heat kernel
$K(t,x)$ of the relevant wave operator. Thus we may compute
the change of $\Gamma$ under conformal changes of ${\M}$ from the
heat kernel expansion alone.\par
To be more specific we consider bosonic and fermionic theories
with classical actions
\begin{equation}
S_B=\int\limits_{\M} \sqrt g\; \Phi\,A(g)\,\Phi,\quad\hbox{and}\quad
S_F=\int\limits_{\M} \sqrt g\; \Phi\,D(e)\,\Phi, \label{2.8}
\end{equation}
where $A(g)$ and $D(e)$ are second and first order
differential operators (e.g. $A(g)=-\Delta_g+\xi\,{\mathcal R}$ for
scalars and $D(e)=\di$ for Dirac fermions). The Weyl invariance
\eqref{2.4} then requires that the wave operators transform as
follows under Weyl transformations of the metric and $d$-bein:
\begin{equation}
A(e^{2\varphi}g_{\mu\nu})=e^{-(\al+d)\varphi}A(g_{\mu\nu})e^{-\al\varphi}
\quad\hbox{and}\quad D(e^\varphi e^a_{\;\mu})=e^{-(\al+d)\varphi}
D(e^a_{\;\mu})e^{-\al\varphi}.\label{2.9}
\end{equation}
Also we assume the matter fields to obey some conformally invariant
boundary conditions on $\RM$.\par
According to \eqref{2.7} the response of the effective action
\begin{equation}
\Gamma[g]=-\log \int {\mathcal D}\Phi \,e^{-S[\Phi,g]}=\pm\ha\log\det\;A
\label{2.10}
\end{equation}
(the plus sign holds for bosons and the minus sign
for fermions for which $A=D^2$)
to a conformal deformation of ${\M}$ is equals to
the difference $\Gamma[e^{2\varphi}g]-\Gamma[g]$.
To determine this difference one introduces the one-parameter family
of metrics
\begin{equation}
g^\tau_{\mu\nu}=e^{2\tau\varphi}g_{\mu\nu}\label{2.11}
\end{equation}
which interpolates between the two metrics, and determines
the $\tau$-variation of the zeta-function regularized determinants
\begin{equation}
\log\det A(g^\tau)=-{d\over ds}\zeta (\tau,s) |_{s=0},\qquad\hbox{where}
\quad \zeta (\tau,s)=\sum\lambda_n^{-s} (\tau ). \label{2.12}
\end{equation}
In what follows we shall assume that all eigenmodes $\Phi_n$ of
$A(g^\tau)$ have positive
eigenvalues $\lambda_n (\tau )$ for $0\leq\tau\leq 1$. Clearly,
the difference of the effective actions (for bosons) is now given by
\begin{equation}
\Gamma[\M,e^{2\varphi}g]-\Gamma[\M,g]=
-{1\over 2}\int\limits_0^1 d\tau\,{d\over ds}{d\over d\tau}\zeta(\tau,s)\vert_{s=0},
\label{2.13}
\end{equation}
and similarly for fermions.
That \eqref{2.12} indeed regularizes the determinants, i.e. the zeta-function
is smooth at the origin, and that the $\tau$-variation of $\Gamma$
can be computed from the heat kernel expansion can be seen
as follows: 
\begin{enumerate} 
\item
First one rewrites the sum defining the zeta-function in \eqref{2.12} as
a Mellin transform of the heat kernel as
\begin{equation}
\zeta (\tau,s)={1\over\Gamma (s)}\int\limits^{\infty }_0 dt\, t^{s-1}
\hbox {Tr}\, e^{-tA(g^\tau)}.\label{2.14}
\end{equation}
The $\tau $-derivative of \eqref{2.14} is obtained if the Hellman-Feynman
theorem
$$
{d\over d\tau}\lambda_n=\Big(\Phi_n,{d \over d\tau}A(g^\tau)\Phi _n \Big)
$$
and \eqref{2.9} with $\varphi$ replaced by $\tau\varphi$ are used,
resulting in a factor $s(2\al+d)$ and an insertion of a Weyl
angle $\varphi $ into the trace of \eqref{2.14}. \par \noindent
\item For sufficiently smooth manifolds the curvature
scalar is bounded and the wave operators of interest are self-adjoint
(e.g. $(-\varDelta _{\tau }+\xi {\mathcal R} _{\tau })$ is self-adjoint
due to a theorem of Kato and Rellich \cite{11}). Using ordinary
spectral theory one shows that
\begin{equation}
e^{-tA(g^\tau)} = -{1\over 2\pi i } \int\limits_{i \infty}
^{-i \infty } e^{-t\lambda } \big(A(g^\tau)-\lambda\big)^{-1} d\lambda \; ,
\label{2.15}
\end{equation}
where the contour encloses the spectrum at infinity. The representation
\eqref{2.15} will be the starting point for our explicit calculations of
the heat kernel expansion in section 4.
\par \noindent
\item For the eigenvalues, there is an estimate \cite{12}
\begin{equation}
\lambda_n > C n^{\delta} \; ,\quad C>0 \; , \; \delta >0 \; ,\label{2.16}
\end{equation}
valid, because the operators under consideration are not only
selfadjoint but also elliptic, a property explained below.
Hence $\int _1 ^{\infty } dt \,t^{s-1} \hbox {Tr}\, \exp(-tA(g^\tau))$
is an entire function of $s$. This suggests to split the integration
region in \eqref{2.14} into $[0,1]$ and $[1,\infty]$.
In the limit $s \rightarrow 0 $ the second integral and its
$\tau $-derivative vanish, and we are left with the first integral.
\par \noindent
\item In order to evaluate the first integral we construct the heat
kernel in the limit $t\rightarrow 0^+ $ up to regular parts.
For the system $(A,\M,\RM )$ it is known that
\begin{equation}
\hbox {Tr }e^{-tA(g^\tau)}\varphi \; \sim
{1\over t^{d \over 2}}\,\sum
_{n=0} ^{\infty } t^{n/2} \Big[\int\limits_{\M }\sqrt{g^\tau}\,
a_{n\over 2}(\varphi,g^\tau_{\mu\nu}) +\int\limits_{\RM }
\sqrt{\tilde g^\tau}\,b_{n\over 2}(\varphi,g^\tau_{\mu\nu})\Big],
\label{2.17}
\end{equation}
where $\tilde g^\tau$ is the determinant of the metric on $\partial\M$
induced by $g^\tau_{\mu\nu}$.
For constant $\varphi$ the Seeley-deWitt coefficients
$a_{n\over 2}$ are local
polynomials in the curvature and its covariant derivatives
and the $b_{n\over 2}$ are local polynomials in the intrinsic and
extrinsic curvatures of the boundaries. The $a$'s vanish for
odd $n$ and have dimensions (length)$^{-n}$. The $b$'s have dimensions
(length)$^{1-n}$.
\end{enumerate}
Using this expansion the $\tau$-derivative of the $\zeta$-function is
defined for $2s>d$ and can be analytically continued, apart
from poles at $s=d/2\;, (d-1)/2\;,...,$ to all
$s$. In the limit $s \rightarrow 0 $ the pole at $s=0$
combines with the asymptotic behaviour of $s/\Gamma(s)\sim s^2 $
such that
\begin{equation}
\delta\Gamma=
-\Big(\al+{d\over 2}\Big)\int\limits_0^1 d\tau \,\Big(\int\limits_{\M }
\sqrt{g^\tau}\,a_{d\over 2}[\varphi,g^\tau_{\mu\nu}]+\int\limits_{\RM }
\sqrt{\tilde g^\tau}\,b_{d\over 2}[\varphi,g^\tau_{\mu\nu}]\Big),
\label{2.18}
\end{equation}
and this formula for the change of the quantum action
will be used in the following. Of course, in general $\Gamma$
depends on the chosen renormalization conditions.
The possible counterterms are just the lower Seeley-deWitt
coefficient-functions and their coefficients are determined
by these renormalization conditions. But contrary to the universal
(scheme-independent) result \eqref{2.18} these additional ambigues terms
are not universal.

The corresponding formula for fermions is obtained similarly
and one obtains $-2$ times the expression on the right hand side
of \eqref{2.18} where $a_{d\over 2}$ and $b_{d\over 2}$ are now the Seeley-deWitt
coefficients of $D^2(e)$.\par
We see that the variation of the free energy under Weyl transformations,
and thus under conformal transformations, is determined by the
$t$-independent terms in the small $t$ expansion \eqref{2.17} of the
(weighted) heat kernel.
Since the Seeley-deWitt coefficients are computed iteratively the
calculation of the relevant coefficients becomes rather involved in four or
more dimensions. So far the coefficients $a_n,\;n\leq 4$, which are
of interest in $8$ or less dimensions, have been
determined \cite{12,13}. The boundary dependent $b$-terms are especially
difficult
to compute and only the $b_n,\;n\leq 2$ are known \cite{9,10}.\par
Note that \eqref{2.18} automatically yields a separation of $\delta\Gamma$ into its
bulk and surface contributions. As we shall see later the individual
bulk and surface terms are not invariant under general coordinate
transformations, only their sum is invariant. This implies that
they are not independent, and that the bulk terms partly determine
the surface terms. Before computing the relevant $b_2 $-term in four
dimensions, we first investigate the consequences of \eqref{2.7} and \eqref{2.18}
in two dimensions.

\section{Finite Size Effects in $2$ Dimensions}\label{sec:finite2}
In recent years the postulate of conformal invariance for critical
models made it possible to identify them quite successfully with
$2$-dimensional Euklidean conformal field theories \cite{14}. Such
theories
are characterized by the central charge $c$ which is determined
by the singular part of the operator product expansion of
the energy momentum tensor. In two dimensions (and for
topologically trivial regions) one can always find coordinates
for which the metric is conformally flat, $g_{\mu\nu}=
e^{2\varphi}\delta_{\mu\nu}$ and thus \eqref{2.18} allows one to calculate
the metric-dependence of the effective action for arbitrary
$2$-dimensional space-times. Thus in two dimension $\Gamma[g_{\mu\nu}]-
\Gamma[\delta_{\mu\nu}]$ is completely determined by the Seeley-deWitt
coefficients $a_1$ and $b_1$ (in other regularizations as the one
chosen here $a_{1\over 2},\,b_{1\over 2}$ and $a_0$ may be needed as
counterterms leading to extra non-universal terms in $\Gamma$). Since
vacuum expectation values of products of the energy momentum tensor
can be computed from the effective action as
\begin{equation}
\langle T_{\mu\nu}(x_1)\cdots T_{\al\beta}(x_n)\rangle={2^n\over
\sqrt{g(x_1)\cdots g(x_n)}}{\delta^n \Gamma[g]\over
\delta g^{\mu\nu}(x_1)\cdots \delta g^{\al\beta}(x_n)},\label{3.1}
\end{equation}
they are determined by the volume part of the effective action alone,
and thus by the coefficient $a_1$. In particular the
central charge is determined by this coefficient.\par
Let us now apply the result \eqref{2.7} to a region in flat Euklidean
space-time for which $g_{\mu\nu}=\delta_{\mu\nu}$ in \eqref{2.7}. In two
dimensions the (global) conformal group $SO(2,2)$
introduced in the previous section
is only a small subgroup of all
conformal transformations of the Euklidean plane since all analytic point
transformations
\begin{equation}
w=w(z)\quad\hbox{and}\quad \bar w=\bar w(\bar z),\qquad z=x^0+ix^1,
\;w=y^0+iy^1\label{3.2a}
\end{equation}
are conformal, and thus the result \eqref{2.7} applies to all of them with
\begin{equation}
e^{2\varphi (z,\bar z)}={dw\over dz}{d\bar w\over d\bar z}\,.\label{3.2b}
\end{equation}
According to the Riemann mapping theorem \cite{15} any region with smooth
boundary (and without hole) can be mapped
into the unit disk by an analytic transformation. Thus the formula \eqref{2.7}
determines the effective actions for arbitrary shaped regions
relative to the effective action for the unit disk.
For example, for $A(g)=-\varDelta_g$ and Dirichlet boundary conditions
the coefficients $a_1$ and $b_1$ are given in appendix B
and the general formula \eqref{2.7} together with 
(\ref{3.2a},\ref{3.2b}) yield
\begin{align}
\Gamma [ {\mathcal N },\delta_{\mu\nu}] - \Gamma [ {\mathcal M},
\delta_{\mu\nu}]
= & - {i \over 48 \pi} \int\limits_{{\mathcal M}} 
\Big({d\over dz} \log{dw \over dz }\Big)
\Big({d\over d\bar z} \log {d\bar w \over d \bar z }\Big)
dz d\bar z \nonumber\\
&- {i \over 48 \pi} \oint\limits_{\pa {\mathcal M }} 
\Big({d\over d \sigma} \log
{z^{\prime} \over \bar z ^{\prime}}\Big) 
\log\Big({dw \over dz} {d\bar w \over d \bar z } \Big) 
d \sigma \; , 
\label{3.3}
\end{align}
where ${\mathcal M}$ is mapped into ${\mathcal N}$ by the conformal transformation
$w=w(z)$. We have used that for scalars $\al+d/2=1$.
The line-integral along the boundary $\pa\M$ of $\M$ contains
the derivative $z^\prime=\pa _{\sigma} z $ of
the parametrised boundary curve $z(\sigma)$
with respect to the curve parameter $\sigma$.
Note that on the Euklidean plane only the surface term in
\eqref{2.18} contributes to the effective action since $a_1$ vanishes,
and indeed the first term on the right hand side of \eqref{3.3} can be converted
into a boundary term.\par
If one considers dilatations, the finite size effects
are independent of the shape of the boundary, simply given by
the Euler number $\chi^E $ of the manifold
\begin{equation}
\Gamma[\lambda\M]-\Gamma[\M]=-{1\over 6}\,\chi ^E \log \lambda ,\qquad
\chi ^E = {1\over 2\pi} \int\limits_{\RM } \trc . \label{3.4}
\end{equation}
Here $\trc $ is the trace of the extrinsic
version of the second fundamental form (see appendix A)
and the sign in the definition of $\chi ^E $ is chosen to be positive for
a sphere.\par
Let us now see how the bulk term $\int a_1$ (which determines
the $T_{\mu\nu}$ correlators) and the surface term $\oint b_1$ (which
determines the finites size effects) are related. For that one observes
from the explicit expressions \eqref{B.1} and \eqref{B.4}, that for a non-zero
${\mathcal R}$ both terms are not seperately invariant under the
transformations \eqref{3.2a} taken as diffeomorphism so that
$$
\hat\varphi(w,\bar w)=\varphi(z(w),\bar z(\bar w))
+{1\over 2}\log \Big({dz\over dw}{d\bar z\over d\bar w}\Big).
$$
Only the sum of the bulk and the surface term is invariant (up to
$\varphi$-independent terms, reflecting the non-invariance of
$\Gamma[\M,\delta_{\mu\nu}]$ under the transformations \eqref{3.2a} and
leading to finite size effects in flat space-time) 
and this fixes the relative normalization of $a_1$ and $b_1$. Thus
the correlators \eqref{3.1} and finite size effects \eqref{3.3} are very
much related. This relation can only be seen when the
scalar field is coupled to a non-trivial background metric.
More generally, for an arbitrary conformal field theory
the coefficient $a_1$ must be a local, dimension $2$ object which
is a scalar for $\varphi=1$. The only such object is the Ricci scalar,
so that $a_1$ must have the form \eqref{B.1}, up to a constant
factor $c$, and $c$ is the central charge. From general covariance we
conclude that $b_1$ must
have the form \eqref{B.4} times the same constant $c$. It follows that
the central charge define via the short distance expansion
of $\langle T_{\mu\nu}(x_1)T_{\al\beta}(x_2)\rangle$
reappears in the formulae (\ref{3.3},\ref{3.4}). 
In particular \cite{16}
\begin{equation}
\Gamma[\lambda\M]-\Gamma[\M]=-{c\over 6}\,\chi ^E \log \lambda ,\label{3.5}
\end{equation}
for a conformal field theory with central charge $c$. Stricly speaking
the volume terms do not determine the surface terms uniquely. But
the ambiguous surface terms must be scalars under analytic
coordinate transformations. The only ambiguous term in $2$ dimensions
is $\int_{\M}\varDelta \varphi$, and such a term does not contribute
on flat space-time.
\def \xia {p _a }
\def \xib {p _b }
\def \xic {p _c }
\def \aip {\partial^{\alpha }_p a_k }
\def \lap {A + {i\lambda \over t }}
\def \kaa {{\mathcal K} _{aa}}
\def \chrbaab {\Gamma ^b_{aa,b}}
\def \chrbaarb {\tilde \Gamma ^b _{aa,rb}}
\def \cchrbaab {\Gamma ^b_{aa,b}}
\def \cchrbaarb {\Gamma ^b _{aa,rb}}
\def \trcc {{\mathcal T}r\chi^2}
\def \trccc {{\mathcal T}r\chi^3}
\def \Ric {{\mathcal R} _{\mu \nu}}
\def \Riem {{\mathcal R}_ {\sigma \mu \rho \nu} }
\def \cmunu {\chi ^{\mu \nu }}
\def \nmu {n^{\mu}}
\def \nnu {n^{\nu}}
\def \nro {n^{\rho}}
\def \nsi {n^{\sigma}}
\def \aphi {a_{{l\over 2}}[\varphi,g_{\mu\nu}]}
\def \bphi {b_{{l\over 2}}[\varphi,g_{\mu\nu}]}
\def \bbbbbphi {b_{2}[\varphi,g_{\mu\nu}]}
\def \bbbbphi {b_{{3\over 2}}[\varphi,g_{\mu\nu}]}
\def \bbbphi {b_{1}[\varphi,g_{\mu\nu}]}
\def \bbphi {b_{{1\over 2}}[\varphi,g_{\mu\nu}]}
\def \aaaaphi {a_{2}[\varphi,g_{\mu\nu}]}
\def \aaaphi {a_{1}[\varphi,g_{\mu\nu}]}
\def \pnmu { n^{\mu } \nabla _{\mu}\varphi }
\def \gchrx {g^{ab} \Gamma ^c _{ab} p _c }
\def \pxi {\partial _p}
\def \pxixi {\partial ^2_p}
\def \poxi {\partial ^p}

\section{Heat Kernel Expansion for Manifolds with Smooth Boundaries}
\label{sec:heat}
In this section we outline the method used
to calculate $b_{2}$. More extensive expositions
may be found in \cite{6} and \cite{13,17,18,19,20}.
In the previous sections we have seen that the singular structure of the
heat kernel trace for small parameters $t$ is required.
Starting from \eqref{2.15}, the techniques of pseudodifferential
operators can be employed to investigate this singular
behaviour. In this formalism two operators are identified if
they possess the same singularity structure.
In particular the inverse operator appearing in \eqref{2.15} is
constructed up to smooth parts. \par
More precisely, an equivalence relation will be defined
\begin{equation}
A\sim B \quad \Leftrightarrow \quad A-B :D \rightarrow
{\mathcal C}^{\infty }  \; ,    \label{4.1}
\end{equation}
where the function space $D$ will be specified later on. Operators
like $A-B$ do not produce singularities and are viewed
as being neglegible. The procedure reminds of Lebesgue theory,
where all results are valid up to sets of measure zero.
\par \noindent
In a first step we consider $R^d$ instead of the general
case $(\M,\RM)$. Let $A(x,D)=\sum _{|\al| \leq m} a_{\al} (x) D^{\al}_x
$ be a differential operator with $C^{\infty}$-coefficient functions
$a_\alpha$. We use the multi-index notations
\begin{align}
\al =(\al_1,......,\al_d )\,,\quad
\vert\al\vert &= \sum_{i=1}^d \al _i\,,\quad
\al! =\al_1! .......\al_d ! \,,
\nonumber\\
D_x^\al &={1 \over i^{|\al |}} {\partial \over \partial _{x_1}^{\al _1} }.....{\partial \over \partial _{x_d}^{\al _d}}\; . 
\label{4.2}
\end{align}
With $A$ we associate a polynomial, called symbol, replacing the
derivatives in $x$ by the momentum $p\in R^d $
\begin{equation}
\si _A(x,p) = \sum _{|\al| \leq m} a_{\al}(x)p ^{\al}  \quad , \quad
\si _A ^m =\sum _{|\al |=m} a_{\al} (x)p ^{\al} \; .\label{4.3}
\end{equation}
$\si _A^m $ is called leading symbol and if it is different from
zero for all non-zero $p $ the operator is called elliptic.
We recover the differential operator $A$ from its symbol $\si _A $
by Fouriertransformation
\begin{equation} 
A u(x) =\int e^{i\langle x,p\rangle }\si_A(x,p ) \hat u (p ) dp,\qquad
\hat u(p) ={1\over (2\pi)^d}\int e^{-i\langle x,p\rangle } u (x ) dx
\;.\label{4.4}
\end{equation}
The calculations simplify considerably when these polynomials
are used instead of the corresponding operators. The prize we
pay is that we must introduce equivalence classes of operators to recover,
for example, the inverse operator from the inverted symbol. The reason is
that the inverted symbol usually has singularities which must be
regularized by introducing a cut off function. In addition, to derive
\eqref{2.17} from \eqref{2.15}, we must scale $\lambda$ as
\begin{equation}
\lambda \rightarrow \; -{i\lambda \over t}\; ,  \label{4.5}
\end{equation}
so that $t$ appears in the inverse $ (A(g)+i\lambda/t)^{-1}$ and
hence in the inverted symbol. Unfortunately the latter lacks a
homogeneity property and $t$ can not be removed from it.
But it turns out that we can find a sequence of homogeneous symbols
approaching the exact inverse in the sense of \eqref{4.1}, if
the function space $D$ is the Schwartz class supplied with the
Sobolev norm, denoted by ${\mathcal H }^s, s \in R $,
$$
\| u \|_s = \biggr (\int \hat u (p ) (1 + |p | ^2)^s dp
\biggl ) ^{{1\over 2}} \; .
$$
These spaces have the following properties:
\begin{enumerate} 
\item $ D_x^\al : {\mathcal H}_s \rightarrow {\mathcal H}_{s-|\al | } $ is
continuous.                  \par \noindent
\item Let $d$ be the dimension of space. For $s>d/2 $
\begin{align*} 
&u \in \H _{s+k} \Rightarrow u \in \C ^k  \quad \hbox {(Sobolev-Lemma)}
\\
&u|_{\M }\rightarrow u|_{\RM }  \quad \hbox {is continuous
and surjectiv}
\; .
\end{align*}
\end{enumerate}
To treat the inverse $(A-\lambda)^{-1}$ let us introduce the notion
of parameter dependent symbols.
$\si \in \S^m $ if
\begin{align}
i) \; \; & \si (x,p,\lambda)\;\hbox{is holomorphic in} \; \lambda \,,
 \nonumber \\
ii) \;\;&
|D_x^{\al}D_{p}^{\beta} D_{\lambda}^{\gamma}\si | \leq C_{\al\beta
\gamma } (1+|p|+|\lambda|^{1/2} )^{m-|\beta|-2\gamma } \; .
\label{4.6}
\end{align} 
We then establish the following equivalence relation on symbols
\begin{align}
\si \sim \si ^{\prime} &\Leftrightarrow
\si-\si ^{\prime} \in \S ^{-\infty } \quad , \quad \S^{-\infty
}:= \bigcap _{m \in {\mathcal R}} \S ^m\nonumber\\
& \Leftrightarrow A-A ^{\prime} : {\mathcal H}_s \rightarrow
C^{\infty} \hbox{ for all } \; s \; .
\label{4.7}
\end{align}
The next step is to define a symbol product
\begin{equation}
\si (AQ) = \sum _{\al }{1\over \al!} \partial^{\al } _{p } \si_A\,
D^{\al}_x \si_Q \; .    \label{4.8}
\end{equation}
Now we are ready to approximate, in the sense of \eqref{4.1}, the inverse
$A^{-1}$ of $A$, ($\si _A \in \S^m $) by an operator $Q$.
The symbol $q$ of $Q$ is obtained as follows:
\begin{enumerate} 
\item Using truncations one proves the existence of a $q$
\begin{equation}
\si _A q\sim q \si _A \sim 1 \; \hbox{ and }\; q \in \S^{-m}\; .\label{4.9a}
\end{equation}
\item  For $q $ one makes an ansatz
\begin{equation}
q = q_{-m}+q_{-m-1}+......\quad ,\quad q_i \in \S^{-i} \;
\label{4.9b}
\end{equation}
where the sum on the right hand side
uniquely determines a symbol. \par \noindent
\item Calculate the $q_i $ iteratively from 
(\ref{4.9a},\ref{4.9b}) by using \eqref{4.8}.
\end{enumerate} 
So far we haven't made any particular choice for the wave operator
$A$. In what follows we shall consider scalar particles
for which this operator has the form
\begin{equation}
A=-\varDelta+\xi{\mathcal R}.\label{4.10}
\end{equation}
Let us now apply the above algorithm to scalar particles for which
$\si_{A+i\lambda/t}$ is the second degree polynomial
\begin{align}
\si_{\lap  } =\; &a_2 + a_1 + a_0
\quad , \quad  a_i \in \S ^i \nonumber\\
a_2 =\; g^{\al \beta} p _{\al} p_\beta &+{i\lambda \over t}, \qquad
a_1 =\;ig^{\al \beta} \Gamma ^{\gamma } _{\al \beta} p _{\gamma},\qquad
a_0 =\; \xi\, {\mathcal R} \; .\label{4.11}
\end{align}
The $a_i$ are homogeneous in the momentum $p$ and $\sqrt{\lambda}$.
It is of main importance to include the $\lambda $ dependent part into
$a_2$. The ansatz for the symbol of the approximating inverse reads
\begin{equation}
q\sim q_{-2}+q_{-3}+q_{-4}+....\,,\qquad q_{-i}\in \S ^{-i}\;.\label{4.12}
\end{equation}
Using (\ref{4.9a},\ref{4.9b}) one finds
\begin{equation}
q_{-2}= {1\over a_2} ,\label{4.13}
\end{equation}
which is nowhere singular and homogeneous in $p $ and $\sqrt{\lambda}$.
Having made this choice the product of \eqref{4.11} with \eqref{4.12}
yields (1+ lower order terms). The lower order terms
are grouped together, each group belonging to $S^{-l}$
for some $l$, and then separately set equal to zero
\begin{equation}
a_2\,q_{-2-l} +\sum _{j<l} {1\over \alpha !} \aip \qp =0\qquad
|\al|+2+j-k=l \quad \forall\,l \geq 0 \; . \label{4.14}
\end{equation}
This algebraic system of equations must be solved for the $q$'s, which
are easily seen to have the homogeneity property
\begin{equation}
q_{-2-l}\,(x,p,i\lam/t)
=t^{1+l/2}\, q_{-2-l}(x,\sqrt{t}\,p,i\lam) \; . \label{4.15}
\end{equation}
By this procedure it is guaranteed, that after infinitely many steps
the remaining contribution is in $\S^{-\infty }$.
Because of homogeneity the substitution
\begin{equation}
p \rightarrow \sqrt{t}\, p \label{4.16}
\end{equation}
allows us to factorize the $t$-dependence in all integrals
to be performed. \par
As explained in section 2, we are need to determine
the heat kernel trace with the insertions of a conformal angle $\varphi$.
Denoting as usual by $\hat \varphi$ the Fourier transformed angle we
arrive at
\begin{align}
\exp [ -t A(g)] \varphi (x)&=\int_ U dz\langle x|
e^{ -t A(g)}|z \rangle \langle z|\varphi \rangle
\label{4.17}\\
&\sim {1\over 2\pi}\sum_l \int  d\lam \int dp\,
e^{ i\lam }e^{ i\langle x,p\rangle } q_{-2-l} \,\hat \varphi (p) \; .
\nonumber 
\end{align}
The integrals are absolutely convergent,
and we may interchange the order of integration. Therefore
\begin{equation}
\langle x|e^{-tA(g)}\varphi|x\rangle \sim
t^{-{d \over 2}}\sum_l \aphi\; t^{{l\over2}}\label{4.18a}
\end{equation}
with
\begin{equation}
\aphi ={1\over (2\pi )^{d+1}}\int\limits_{{\mathcal R}^d} dp
\int\limits_{-\infty}^{\infty} d\lam \,e^{i\lam}  q_{-2-l} \;\varphi
\; .  \label{4.18b}
\end{equation}
The notation $\aphi $ is used in accordance with the existing
literature on the subject.\par
To generalize from $R^d$ to $\cal M$ the following remarks are
in order: Using a decomposition of one on ${\M}$, compatible with an atlas,
one can derive the results (\ref{4.18a},\ref{4.18b}) in each chart. It follows that the trace
in \eqref{2.17} is a sum of the contributions from the different charts
covering space-time. It can be shown that the result is independent
of the chosen atlas. \par
We now turn to the general case with boundary.
Near the boundary we introduce geodesic coordinates $x^\al =(r,x^a)$,
$\al,\beta,..=0,1,2,\dots,d-1,\;\; a,b,..=1,2,\dots,d-1$.
$x^a$ is the point on the boundary minimizing the geodesic distance
to $x^{\al }$ and $r$ is the geodesic distance.
Also we use Riemann normal coordinates on the boundary.
This way we may identify a neighbourhood of a point on the boundary with
a region in $(R,R ^{d-1})$, the boundary being given by $r=0$. Points
in ${\mathcal M}$ have then positive $r$ and those in $X\setminus M$ have
negative $r$. Again using a decomposition
of one we may even assume this region to be $(R,R^{d-1})$.
In these coordinates the metric has the form
\begin{equation}
g_{rr}=\;g^{rr}=1,\qquad g_{ra}=\;g^{ra}=0,\qquad ds^2=\;dr^2 +
g_{ab} (r,x)\,dx^a dx^b  \label{4.19}
\end{equation}
and then the intrinsic version of the second fundamental form
(see appendix A) simplifies to
$$
\kab =\Gamma ^r_{ab}=-{1 \over 2} \gabr.
$$
At the origin we have
\begin{equation}
\gab |_0 =\delta _{ab},\qquad \chr |_0 =0,\qquad \trk |_0 =\k _{aa}|_0
 \; .  \label{4.20}
\end{equation}
In what follows we denote the restriction of $g_{ab},\;g_{ab,r}$ etc.
to $\partial{\M}$ as
$$ \gab|_{r=0}=: \ti g_{ab},\qquad \gabr|_{r=0}=:\tilde g_{ab,r}\qquad
\hbox{etc.}
$$
Now we would like to generalize (\ref{4.18a},\ref{4.18b}) when boundaries are present.
We supplement \eqref{4.10} with the conformally invariant Dirichlet boundary
conditions
\begin{equation}
\Phi |_{\RM } =0  \label{4.21}
\end{equation}
for $\Phi $ in some Sobolev space. It follows that
\begin{equation}
(A+i\lambda )^{-1} \Phi |_{\RM } =0 \; .\label{4.22}
\end{equation}
Now we define restriction and extension operators $\hat r$,$\hat e$ as
follows: For $\Phi$ defined on $\M $
\begin{equation}
\hat e \Phi : =\begin{cases} \Phi  &\text{if }\;r\geq 0\,, \\
                      0 &\text{if }\;r<0\,, 
                      \end{cases}\label{4.23}
\end{equation} 
and $\hat r $ restricts functions on $X$ to $\M $. Assume that we
have already constructed an approximative inverse $Q$ of
$A+i\lambda$ on $X$ (as explained above). Its restriction $\hat r Q \hat e $
does not obey to the boundary condition and hence
\begin{equation}
G_{\lambda }\;:\;=\Big( \hat rQ\hat e
-{1 \over A+i\lambda} \Big) \; . \label{4.24}
\end{equation}
does not vanish in the sense of \eqref{4.1}. However we can use the previous
results for the boundary-less case to calculate this correction. It is
determined by
\begin{equation}
(A+i\lambda )G_{\lambda } =\;0 \qquad\hbox{and}\qquad
G_{\lambda } |_{\RM } =\; \hat rQ\hat e\;|_{\RM } \; .\label{4.25}
\end{equation}
It can be shown, that $G_{\lambda }$ is
only relevant near the boundary \cite{14}, because if $
a,b$ are truncations having arbitrarily small support around $\RM $ then
the last two terms in
\begin{equation}
G_{\lambda }= a G_{\lambda }b + (1-a) G_{\lambda }
+ a G_{\lambda }(1-b) \; . \label{4.26}
\end{equation}
have $C^{\infty } $ kernels and are
neglegible. Thus the system \eqref{4.25} need only be solved on
$(R^{d-1},R^+)$ near $r=0$. In addition to \eqref{4.25} one must demand that
\begin{equation}
G_{\lambda }|_{r\rightarrow \infty}=0 \label{4.27}
\end{equation}
which is one of the reasons why we can ignore the finite sizes of charts
\cite{6}.\par
To handle the general case we introduce boundary symbols $\ti\si$. They
are polynomials in $ p ^a $ but remain differential operators in
$r$. For scalar particles
\begin{equation}
\lap =-\prr - \ {1\over 2} \ggr \pr -
\gabpp + g^{ab}\Gamma ^c_{ab}\partial _c
+\al \R + {i\lambda \over t} \label{4.28}
\end{equation}
has the full symbol
\begin{align}
\si_{\lap }& = a_2 + a_1 + a_0\nonumber\\
a_2=\t ^2+\ro ^2,\qquad
a_1=&-\frac{i}{2}\ggr\t+i\gchrx \qquad a_0 = \xi\,{\mathcal R}\;.
\label{4.29}
\end{align}
where we have introduced
\begin{equation}
\ro^2 := \goab p_a p_b + {i \lambda \over t}   \label{4.30}
\end{equation}
and $(\tau,p^a)$ are the momenta conjugate to $(r,x^a)$.
Because of \eqref{4.26} it is natural to expand the boundary
symbol around $r=0$.
\begin{align}
\ti \si (x,r,p , D_r ,\lam ) =&\sum _{k,l} r^k \pr ^k a _l(x,0,p
,D_r )=\sum _l \ti a^{(l)} \nonumber\\ 
\ti a^{(l)} =&\sum _{j-k=l} r^k \pr ^k a_j /k! \; .
\label{4.31}
\end{align}
In \eqref{4.31} terms of equal homogeneity are grouped together. 
$\ti a^{(l)} $
has homogeneity $-(m+l) $ in $( 1/r,p ,D_r,\lam ^{1/m} ) $
, if $\ti \si $ is a symbol of order $(-m)$. For scalar particles
the boundary symbols read explicitly:
\begin{align}
\ti a^{(2)} =\;&-\prr + \tgoab \xia \xib +i \lambda    \nonumber\\
\ti a^{(1)}  =\;& r \tgoabr \xia \xib -\ha
\tgoab \tilde g_{ab,r}  \pr + i \tgoab \tchr  \xic \nonumber\\ 
\ti a^{(0)} =\;& \ha r^2 \tgoabrr \xia \xib
-\ha r \tgoabr \tilde g_{ab,r} \pr + \tgoab \tgabrr \pr \nonumber\\
&+ir (\tgoabr \tchr + \tgoab \tchrr )\xic
+\si \xi \tilde \R   \nonumber \\ 
\ti a^{(-1)} = \;&{1\over 6 } r^3 \tgabrrr \xia \xib
-{1\over 4} r^2 \tgoabrr \tilde g_{ab,r} \pr - {1\over 4}
r^2 \tgoab \tgabrrr \pr  \nonumber\\ 
& -\ha r^2 \tgoabr \tgabrr \pr
+ i \ha r^2 \tgoabrr \tchr \xic + i r^2 
\tgoabr \tchrr  \xic   \nonumber\\
&+\frac{i}{2}\, \tgoab \tchrrr \xic + r\si \xi \tilde\R _{,r}
\,.\label{4.32}
\end{align}
For the symbol of $G_{\lambda }$ the ansatz reads
$$
\si_{ G_{\lam }} \sim  \sum d_{-2-l}
$$
leading with \eqref{4.25} to
\begin{equation}
\ti \si _{\lap  } \sum d_{-2-l} =0   \label{4.33}
\end{equation}
and to the boundary conditions
\begin{equation}
d_{-2-l}|_{r=0}=
q_{-2-l}|_{r=0}\qquad\hbox{and}\qquad d_{-2-l} \rightarrow 0
\quad \hbox {for} \quad r\rightarrow \infty  \; .  \label{4.34}
\end{equation}
Using \eqref{4.31} the condition \eqref{4.33} can be rewritten as
\begin{equation}
\ti a^{(2)}d_{-2-l} +\sum _{j<l} {1\over \al !}
\partial^\alpha_p\ti a^{(k)}\ddp =0,\qquad
|\al |+2+j-k=l \; \qquad \forall \,l \geq 0 \; ,\label{4.35}
\end{equation}
which is a system of ordinary differential equations for the $d$'s.
The solutions, subject to the boundary conditions 
\eqref{4.34}, are now
inserted into
\begin{equation}
{1\over 2\pi }\sum_l \int\limits_{-\infty } ^{\infty}
d\lam \int\limits_{R^{d-1}} dp
\int\limits ^{\infty}_{-\infty} d\tau\,
e^{ i\lam }e^{ i\langle x,p\rangle }
d_{-2-l} \;\hat\varphi (p)  \; .\label{4.37}
\end{equation}
There is no exponential factor related to the $\tau $-integration
because the symbols are taken at $r=0$.
Inserting
$$
\hat \varphi =(2\pi)^{-d}\int dzds\,e^{- i\langle s,\tau \rangle }
e^{- i\langle z,p \rangle } \varphi
$$
an exponential factor
$e^{- i\langle r,\tau\rangle }$
reappears in the limit
$x=z$, $r=s$. Scaling the variables according to
\begin{equation}
p'=\sqrt{t}\,p,\qquad \tau'=\sqrt{t}\,\tau\qquad r'={1\over \sqrt{t}} r
\label{4.38}
\end{equation}
and expanding the conformal angle into a Taylor series
$$
\varphi (x,t^{1\over 2}r) =
\sum _0 ^{\infty} {r^n \over n!} \partial _r ^n  \varphi \Big |_{r=0}
\; t^{n\over 2}
$$
the $t$-dependence factorizes again in the occuring integrals.
The result is then
\begin{equation}
G_{\lambda}\sim  t^{-{d \over 2}}\sum _l \bphi
\; t^{{l\over 2}}    \label{4.39a}
\end{equation}
and
\begin{equation}
\bphi = -{1\over (2\pi )^{d+1}}
\sum _{n+k=l-1}\int\limits_{R^{d-1}}dp \int\limits^{\infty }_{-\infty}
d\tau \int\limits_0 ^{\infty }dr \int\limits^{\infty }_{-\infty} d\lam
\cdot e^{i\lam } e^ {-i\langle r,\tau \rangle }
d_{-2-k} (\partial _r ^n \varphi) \Big |_{r=0} \; .\label{4.39b}
\end{equation}
The results of section 2 combined with 
(\ref{4.18a},\ref{4.18b}), (\ref{4.39a},\ref{4.39b}) show that
the relevant Seeley-de Witt coefficients in $d$ dimensions are
those with $l=d$, that is
\begin{equation}
\aphi \qquad\hbox{and}\qquad \; \bphi \; .\label{4.40}
\end{equation}
Taking into account that
$$
a_{{1\over 2}(2l+1)}[\varphi,g_{\mu\nu}]=0
\qquad\hbox{and}\qquad \bphi=0\;\hbox{ for $\RM=0$}\,,
$$
the coefficients relevant for the zeta-function regularisation are
listed in Table \ref{tab:list}.
\begin{table}[h!]
\centering
\begin{tabular}{c|cc}\toprule
dimension&\multicolumn{2}{c}{coefficients}\\
&$\aphi$ & $\qquad\bphi $ \\ \midrule
1&&$\qquad\bbphi$ \\ 
2&$\aaaphi$ & $\qquad\bbbphi$\\
3&&$\qquad\bbbbphi$\\
4&$\aaaaphi$ & $\qquad\bbbbbphi$\\ \bottomrule
\end{tabular}
\caption{Coefficients relevant for the zeta-function regularisation.}
\label{tab:list}
\end{table}

\noindent
A full list of these coefficients is given in appendix B.
To calculate $b_{2} $, we need $d_{-5}$ and solve \eqref{4.35}
successively up to third order. To satisfy \eqref{4.34}, we also need
$q_{-5} $, which is the solution of the algebraic system
\begin{align}
a_2 q_{-2-l} + &\sum _{j<l} (\pr ^{\al } a_k
D^{\al }_r q_{-2-j} + \pxi ^{\al } a_k
D^{\al }_x q_{-2-j} )/\al ! =0  \nonumber\\ 
& k-|\al | -2-j=-l  \quad \forall l \geq 0 \; . \label{4.41}
\end{align}
According to \eqref{4.39b} we must integrate
\begin{equation}
d_{-5}\varphi \; , \;
d_{-4} \;\partial _r \varphi \; , \;
d_{-3} \;\partial _r ^2 \varphi \; , \;
d_{-2} \;\partial _r ^3 \varphi \; . \label{4.42}
\end{equation}
to determine $\bbbbbphi $. Because of \eqref{4.35} the first one is
determined by (recall $\ti\rho=\rho (r=0)$)
\begin{align}
(\pr -\ti \ro )(\pr +\ti \ro ) \d =\; &
{1\over i} \poxi _a \ti a^{(2)} \pox _a d_{-4} + \ti a^{(1)}d_{-4}
-\ha \poxi _{ab} \ti a^{(2)} \pox _{ab} d_{-3}  \nonumber\\
&+{1\over i} \poxi _a \ti a^{(1)} \pox _a d_{-3} + \ti a^{(0)} d_{-3}
-{1\over 6i} \poxi _{abc}\ti a^{(2)} \pox _{abc} d_{-2}
\nonumber\\ 
&- \ha \poxi _{ab} \ti a^{(1)} \pox _{ab} d_{-2}
+{1\over i} \poxi _a \ti a^{(0)} \pox _a d_{-2} + \ti a^{(-1)}
d_{-2}   \nonumber \\
=\; :& \;(A + Br  + Cr^2 + Dr^3  + Er^4
+ Fr^5 )\ex\,, \label{4.43} 
\end{align} 
and has the solution
\begin{align}
\d = \;&\ti q_{-5} \ex + (ar + br^2 + cr^3 + dr^4 +
er^5 + fr^6 ) \ex   \nonumber\\ 
a=\;& -{15\over {8\ti \ro ^6} }F -{3\over {4 \ti \ro ^5}} E
-{3\over {8 \ti \ro ^4} }D-{1\over {4\ti \ro ^3}} C
-{1\over {4\ti \ro ^2 }} B -{1\over {2\ti \ro }} A  \nonumber\\
b=\;& -{15\over {8\ti \ro ^5 }}F -{3\over {4 \ti \ro ^4}} E
-{3\over {8 \ti \ro ^3} }D -{1\over {4\ti \ro ^2}} C
-{1\over 4\ti \ro  } B\nonumber\\
c=\;& -{5\over {4\ti \ro ^4 }}F -{1\over {2 \ti \ro ^3}} E
-{1\over {4 \ti \ro ^2 }}D -{1\over {6\ti \ro }} C  \nonumber \\
d=\;& -{5\over {8\ti \ro ^3 }}F -{1\over {4 \ti \ro ^2}} E
-{1\over {8 \ti \ro }}D  \nonumber\\
e=\;& -{1\over {4\ti \ro ^2 }}F -{1\over {10 \ti \ro }} E\nonumber\\
f=\;& -{1\over {12\ti \ro }}F      \; .\label{4.44}
\end{align}
After that we perform subsequently the integrations
\begin{equation}
\int\limits^{\infty }_{-\infty } d\tau\quad,\quad
\int\limits^{\infty } _0 dr\quad,\quad\int\limits^{\infty} _{-\infty}
e^{i\lam } d\lam \quad,\quad\int\limits^{\infty }_{-\infty }
d^{(3)}\,p\label{4.45}
\end{equation}
and use the formula \cite{21}
\begin{equation}
\int\limits^{\infty } _{-\infty } d\tau\, e^{-ir\tau }
{\tau^m \over (\tau ^2 + \ti \ro ^2 )^{n+1} }
= {(-i)^m (-1)^n \over n! } \pi \partial _{\ti \ro
^2 } ^n (\ti \ro ^{m-1} e^{-\ti \ro r })
. \label{4.46}
\end{equation}
It is easily seen that terms of odd order in $p $ vanish
after integration. This explains why
the coefficients $a_n$, $n$ odd, do not appear in the table.
After the step \eqref{4.35} has been performed the computation can be done
at the origin of the Riemann normal coordinte system. The
corresponding contribution to $b_2[\varphi=1,g]$ is
\begin{align}
\pi ^3 \Big( & -\ft{11}{90}  \kbb \ggoaabb +\ft{1}{9}  \kaa \ggoaabb
-\ft{1}{45}  \kaa \ggoabab + \ft{1}{18} \ggoaabb {\trk } \nonumber\\
&-\ft{2}{9}  \ggoabab {\trk }
 +\ft{1}{3} \cchrbaarb +\ft{2}{3}  \kaa \cchrbaab -\ft{1}{3}  \chrbaab {\trk }  \nonumber\\
&-\ft{1}{20} \ggoaarbb +\ft{7}{30} \ggoabrab
 -\ft{1}{6} \ggaarrr -\ft{1}{15} \ggoaarrr  \nonumber\\
&-\ft{2}{3}  \kaa \ggaarr +\ft{9}{15}  \kaa \ggoaarr
 +\ft{1}{3}  \ggaarr {\trk } + \ft{2}{15} \ggoaarr {\trk }  \nonumber \\
&+\ft{19}{189} ({\trk } )^3
-\ft{109}{63} {\trk }{\trkk }+ \ft{100}{189} {\trkkk }
 -\ft{2}{3} \al  \R _{,r}   \nonumber\\
& +\ft{2}{3} \al  \R {\trk }  \Big)\,,\label{4.47} 
\end{align}
where all functions are to be evaluated at the origin $x^\al=0$
and equal indices are summed over. This expression
can be written covariantly by using the extrinsic version $\chi
_{\mu \nu }$ of the second fundamental form (see appendix A) as follows
\begin{align}
\eqref{4.47}={ 1 \over {(4\pi)^2} } {1\over 1890} \Big(
&80 \trccc -66 \trc \trcc +10 (\trc )^3  \nonumber\\
&105\, \R \trc - 21 \Ric \cmunu  - \ft{189}{2} \nmu \nabla _{\mu } \R \nonumber\\
& -21\, \Ric \nmu \nnu \trc
 +84\,\Riem \nsi \nro \cmunu \nonumber\\
&-\ft{63}{2} \varDelta ^{(3)} \trc
 + 630\, \xi \nmu \nabla _{\mu } \R  \nonumber\\
&-630\, \xi \trc \R  \Big)\;.\label{4.48}
\end{align}
Here $ n^{\mu } $ is the inward pointing normal.
For the remaining terms in \eqref{4.42} we find
\begin{align}
-2\pi ^3 \Bigl (
& \ft{1}{2} \xi \R - \ft{1}{14} (\trc )^2 + \ft{3}{28} \trcc
 + \ft{1}{12} \Ric \nmu \nnu - \ft{1}{12} \R \Bigr ) \pr \varphi 
\nonumber\\
-2\pi ^3 \Bigl (&\ft{3}{20} \trc \prr \varphi - \ft{1}{12} \prrr
\varphi \Bigr ) \label{4.49}
\end{align}
which can be covariantly written as
\begin{align}
{1\over {(4\pi)^2} } {1\over 1890}\, \Big(
&45 \trcc \pnmu + 126\, \trc \tri -9 (\trc )^2 \pnmu\nonumber\\
&-\ft{315}{2} \R \pnmu -\ft{315}{2} n^{\mu } \nabla _{\mu } \tri 
\nonumber\\
&-126\, \trc \trii + 945\, \xi \R \pnmu  \Big)   \; .
\label{4.50}
\end{align}
Using
\begin{align}
&\int _{\RM }\;
\varDelta ^{(3)} \nmu \nabla _{\mu} \varphi  =0  \nonumber\\
&\int _{\RM }\;
\varphi \varDelta ^{(3)} \trc  =
\int _{\RM }\;
\trc \trii  \label{4.51}\\
&\int _{\RM }\;
\nabla _{\mu} ^{(3)}\varphi \nabla _{\nu} ^{(3)}\chi^{\mu \nu} =
-\int _{\RM }\;
\chi ^{\mu \nu} \nabla _{\mu}^{(3)}  \nabla _{\nu}^{(3)} \varphi
 \; \nonumber
\end{align}
with an upper index $\phantom {}^{(3)} $ indicating derivatives
inside the boundary,
the final form that coincides with \cite{8,9} is
\begin{align}
b_2[\varphi,g_{\mu\nu}]
={1 \over (4\pi)^2 } {1\over 1890} \bigg(
\Big[&80 \trccc -66 \trc \trcc +10 (\trc )^3  \nonumber\\
&\;\;+105 \R \trc - 21 \Ric \cmunu  - \ft{189}{2} \nmu \nabla _{\mu } \R \nonumber\\
& \;\;-21 \Ric \nmu \nnu \trc
 +84 \Riem \nsi \nro \cmunu   \nonumber\\
&\;\;+ 630 \xi \nmu \nabla _{\mu } \R
-630 \xi \trc \R   \Big] \varphi \nonumber\\
&+45 \trcc \pnmu + 126 \trc \tri -9 (\trc )^2 \pnmu  &\nonumber\\
&-\ft{315}{2} \R \pnmu -\ft{315}{2} n^{\mu } \nabla _{\mu } \tri 
+ 945 \xi\, \R \pnmu  \bigg) \; .
\label{4.52}
\end{align}
As always, explicit formulae are with respect to
a chosen convention. To allow for a easy comparison of
our results with those in the literature we specified
our conventions in appendix A.
Finally note that after using 
$$
f_{;nn}=\varDelta^{(4)}f-\varDelta^{(3)}f+\trc \,f_{,n},
$$
$R_{nnnn}=0$ and
\begin{align*}
R_{abcb}L_{ac}&=R_{\mu\nu}\chi^{\mu\nu}-R_{anbn}L_{ab}\nonumber\\
&=R_{\mu\nu}\chi^{\mu\nu}-R_{\sigma\mu\rho\nu}n^\sigma n^\rho \chi^{\mu\nu},
\end{align*}
the coefficient $b_2$ becomes identical to the one of Branson and
Gilkey \cite{9}.
\def \troc {{{\mathcal T}r} \chi}
\def \trocc {{{\mathcal T}r }\chi^2}
\def \troccc {{{\mathcal T}r} \chi^3}
\def \g { \sqrt {\tilde g }}
\def \nf { \partial_n\varphi}
\def \fmunu {\varphi _{,\mu \nu }}
\def \fmu {\varphi _{,\mu  }}
\def \fnu {\varphi _{,\nu  }}

\section{Applications to Simple Geometries}\label{sec:applications}
After having derived the explicit form of the heat kernel
coefficient $b_2$ we can now apply
the general formula \eqref{2.18} to determine the change of $\Gamma$
under conformal transformation of an arbitrary
region $\{\M,\partial \M\}$ in $4$-dimensional (flat) Euklidean
space-time. As we shall see the influence of a wall $\partial \M$
on the vacuum fluctuations is more subtle than in $2$ dimensions.\par
Note that for a Weyl angle belonging to
a diffeomorphism the volume integral in \eqref{2.18} vanishes
and the conformal anomaly is solely a surface effect.
This is of course true in arbitrary dimensions as long as the
imbedding space-time $X$ is flat.
To evaluate the surface term in \eqref{2.18} we still must express the curvature
terms in \eqref{4.52} as functions of $\tau \varphi $ (recall that
$g^\tau_{\mu\nu}=e^{2\tau\varphi}\delta_{\mu\nu}$) and perform the
$\tau $-integration. The final result for massless scalars is
\begin{align}
\delta \Gamma=
-{1 \over {(4\pi)^2}}  {1\over 1890} \int\nolimits_{\RM } \g \;
&\Big\{
\Big[ 80 \troccc -66 \troc \trocc +10(\troc )^3  \nonumber\\
&\quad +21 (\troc )^2 \nf -21 \trocc \nf \nonumber\\
&\quad-14 \troc \nf^2 -21  \cmunu \fmunu
+14  \cmunu \fmu \fnu \nonumber\\
&\quad +21 \troc \tri -28 \tri \nf
+28 \nmu \nnu \fmunu \nf \nonumber\\
&\quad -\ft{21}{2} \fnu \varphi ^{,\nu } \nf
-21 \troc \fmunu \nmu \nnu
\Big] \varphi \nonumber\\
& + 45 \trocc \nf - 18 \troc (\nf )^2
+18 (\nf )^3 \nonumber\\
& -9 (\troc )^2 \nf  +126 \troc \tri
+126 \troc \fmu \varphi ^{,\mu } \nonumber\\
& +126 \tri \nf + 126\, \fmu \varphi ^{,\mu } \nf
-\ft{315}{2} \partial_n \tri \nonumber\\
& -\ft{315}{2} \partial_n (\fnu \varphi ^{,\nu } )\Big\}
\; ,\label{5.1}
\end{align}
where $\partial_n=n^\al\partial_\al$ is the (inward) normal
derivative and all contractions in traces and derivatives are
understood with respect to the original undeformed metric
$g^0_{\mu\nu}=\delta_{\mu\nu}$. Of course $\tilde g$ is the
determinant of the metric on $\partial\M$ induced by
$g^0_{\mu\nu}$.
\subsubsection*{Dilatations}
Now we generalize the two-dimensional result \eqref{3.4} for the change of
$\Gamma$ under dilatations \eqref{2.3a} to four dimensions. Since then
the Weyl angle is constant, $\varphi=\log\lambda$, all but the
first three terms in \eqref{5.1} vanish so that
\begin{equation}
\Gamma[\lambda\M]-\Gamma[\M]=
-{1 \over {(4\pi)^2}}  {\log\lambda \over 1890} \int\limits_{\RM }
\g\; \Big[ 80\,\troccc -66\,\troc \trocc +10\,(\troc )^3 \Big]\; .
\label{5.2}
\end{equation}
Contrary to the two-dimensional case the right hand side
is not a topological invariant. To see that more clearly
let us introduce the Euler number $\chi ^E $.
Applying the index theorem to the
De Rham-complex \cite{22} the Euler number of a bounded flat manifold is
\begin{equation}
\chi ^E = -{1 \over 12\pi^2} \int\limits_{\RM } \g \;
\Big[ 2 \troccc -3 \troc \trocc +(\troc )^3 \Big].
\label{5.3}
\end{equation}
It is just the winding number of the normal map $n:\partial\M
\to S^3$ and the sign convention is such, that it is one for
a sphere. Thus \eqref{5.2} can be written as
\begin{equation}
\delta \Gamma=-{\log\lambda\over 180}\;\chi^E
-{\log\lambda \over 280\pi^2}\int\limits_{\RM } \g \;f(\chi)\label{5.4a}
\end{equation}
where we have introduced the third order polynomial
\begin{equation}
f(\chi )=\;\troccc -\troc\,\trocc +{2\over 9}(\troc )^3  \; .\label{5.4b}
\end{equation}
Contrary to the Euler number the last term in \eqref{5.4a} depends on
the shape of the surface and thus is not topological.
For simple geometries we obtain:
\par \noindent
{\it Spherical bubbles}: For a spherical bubble, $\M={\mathcal B}$,
the boundary surface is a $3$-sphere for which $f(\chi)=0$. It follows that
\begin{equation}
\Gamma[\lambda{\mathcal B}]-\Gamma[{\mathcal B}]=-{1 \over 180 } \log\lambda
\label{5.5}
\end{equation}
leading to a repulsive Casimir force as in
$2$ dimensions (see \ref{3.4}). \par
\begin{figure}[h]
	\centering
	\includegraphics[scale=0.9]{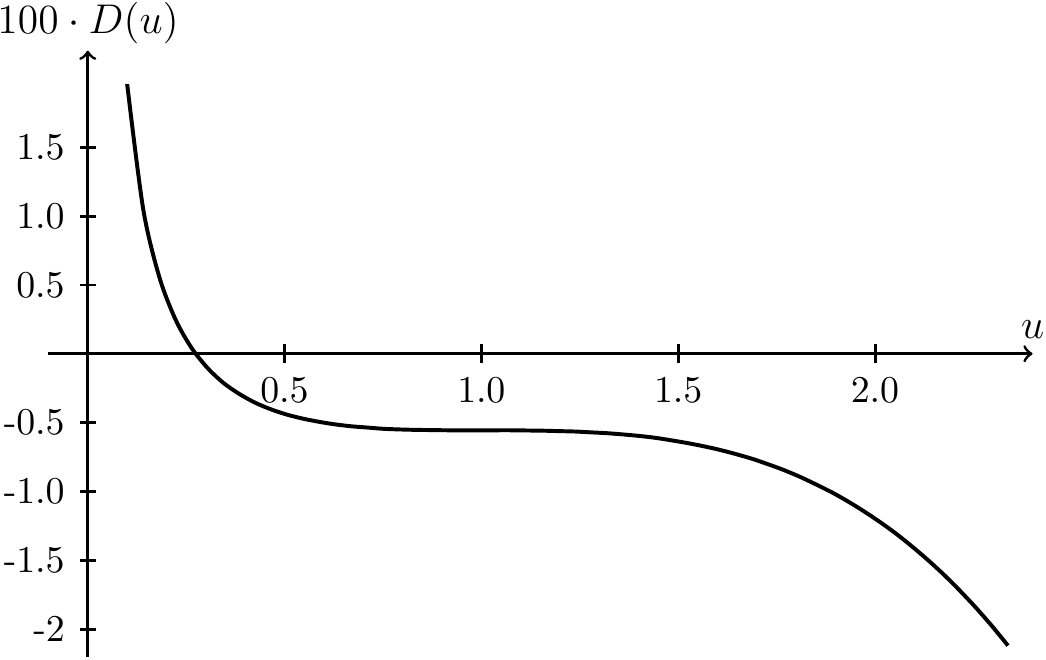}
	\caption{Change of the quantum actions of squashed or stretched bubbles
		under dilatations as a function of the bubble excentricity $u$.
		A stretched bubble has $u<1$ and a squashed one $u>1$.
	}
	\label{fig1}
\end{figure}

 \noindent
{\it Squashed and stretched bubbles}: We parametrize the surface of the
ellipsoid as
\begin{equation} \{x^0,x^1,x^2\}=A \sin\al\;\{\sin\beta \cos\gamma,
\sin\beta\sin\gamma,\cos\beta\},\quad x^3= B\cos\al\;, \label{5.6}
\end{equation}
where
$$
0 \leq \gamma \leq 2 \pi \quad\hbox{and}\quad 0 \leq \al ,\beta \leq \pi.
$$
Inserting the corresponding second fundamental form into $f(\chi)$
in ()\ref{5.4a},\ref{5.4b}) yields
\begin{equation}
\delta \Gamma=\log\lambda \;D(u),\qquad
D(u)=-\Big[{(u-1)^3 (5u^3 + 20 u^2 +29 u + 16) \over
5040 u (1+u) } + {1 \over 180 } \Big] , \label{5.7}
\end{equation}
where we have introduced the parameter $u=A/B$ which measures
the deviation from a spherical bubble. The graph of $D(u)$ is displayed in
Figure~\ref{fig1}. 

One sees that the quantum action increases or decreases
with the volume depending on whether $u <0.274$ or $u>0.274$.
Thus, the vaccum fluctuations try to shrink bubbles stretched in
the $x^3$-direction and expand squashed bubbles.
\subsubsection*{Special conformal transformations}
\begin{figure}[h]
	\centering
	\includegraphics[scale=0.9]{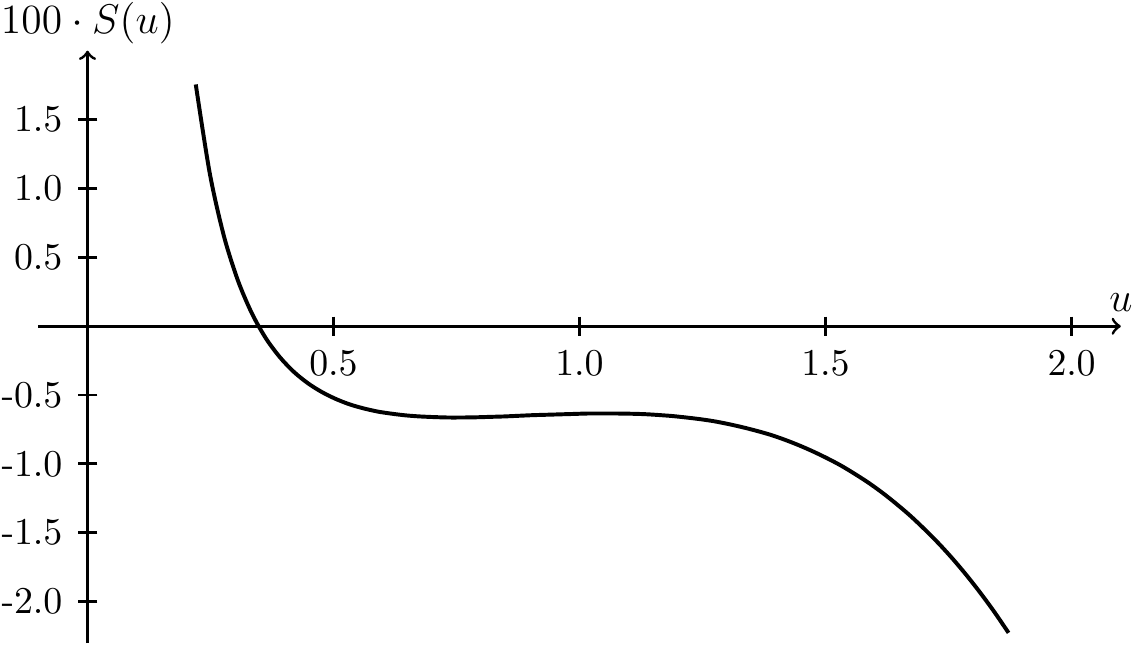}
	\caption{
		Change of the quantum actions of squashed or stretched bubbles
		under special conformal transformations. Depending on the
		excentricity the bubble favours or resists a conformal deformation.	}
	\label{fig2}
\end{figure}

\begin{SCfigure}[1.3][h!]
		\hskip20mm\includegraphics[scale=0.9]{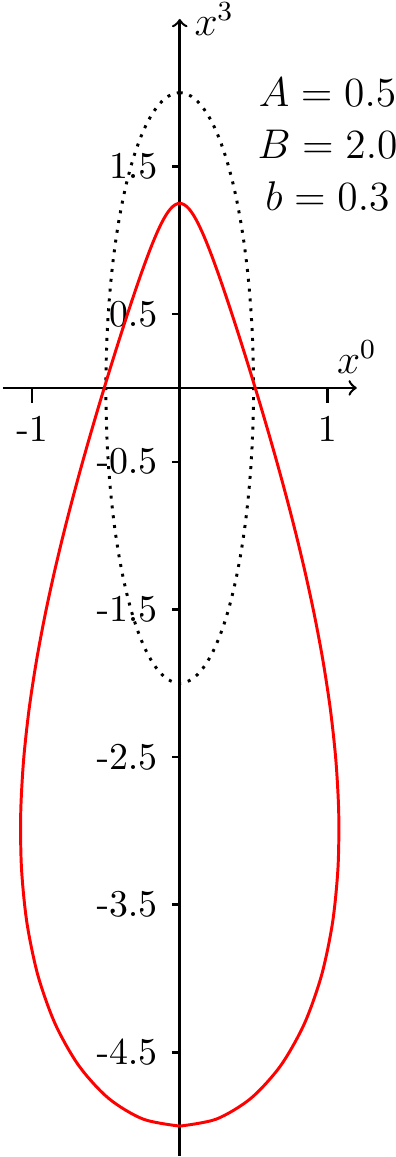}
	\caption{
	The stretched bubble enclosed by the broken line is mapped
	into the body enclosed by the full line by a special conformal
	transformations. The quantum system resists this deformation.\label{fig3}}		
\end{SCfigure}

\begin{SCfigure}[0.6][h!]
	\hskip1mm\includegraphics[scale=0.9]{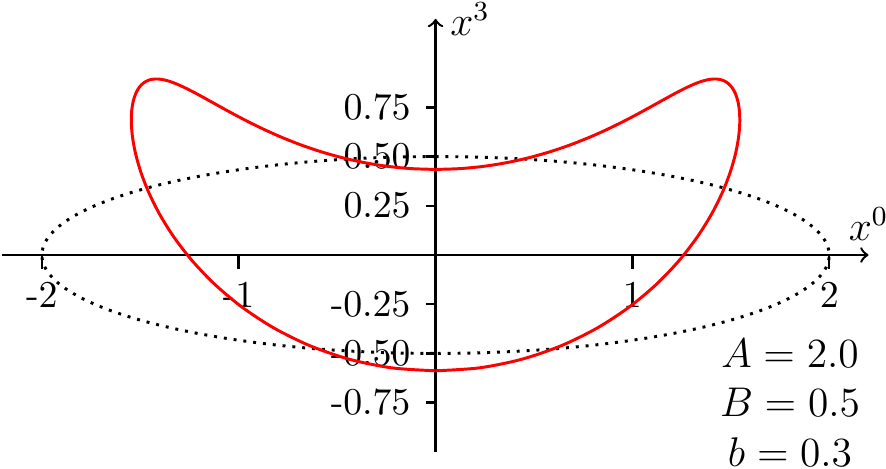}
	\caption{
		The squashed bubble enclosed by the broken line is mapped
		into the body enclosed by the full line. This deformation
		decreases the quantum action.	\label{fig4}}
\end{SCfigure}

A special conformal transformations \eqref{2.3b}
is a composition of an inversion, translation and again an inversion.
We require that bounded regions are transformed into
bounded ones which means that the transformed regions should
contain the origin of the coordinate system if and only if the
original body contained it. For an ellipsoid \eqref{5.6} and a special
conformal transformation \eqref{2.3b} with $b^\mu=(0,0,0,b)$ this
is fulfilled for $Ab,Bb< 1$. Thus we may expand \eqref{5.1} (for
special conformal transformations $\varphi$ is $x$-dependent) in
$Ab$ and $Bb$, and the first non-vanishing terms are of second order.
The explicit result up to second order reads
\begin{equation}
\delta \Gamma=(Ab)(Bb)\;S(u)\label{5.8}
\end{equation}
where
\begin{equation}
S(u)=-{5u^9+10u^8+6u^7+2u^6-32u^5-66u^4+182u^3
+38u^2-9u-8\over 5040\,u^2 (1+u)^2}.\label{5.9}
\end{equation}
The function $S(u)$ is displayed in Figure~\ref{fig2}. 
Again
one finds that stretched bubbles resist a deformation and squashed
ones are unstable against the special conformal deformations \eqref{2.3b}.
Bubbles with $u=u_c=0.348$ are marginal in the sense that their
deformation does not change the quantum action induced by the
vacuum fluctuations.
The figures 3 shows two typical deformation of ellipsoids (the figures
show the intersection of the ellipsoid with the $(x^1,x^2)$-plane). In 
Figure~\ref{fig3} a stretched bubble with $(A,B)=(0.5,2)$, drawn with broken line,
is deformed with $b=0.3$ into the body drawn with the unbroken
line. This deformation increases the quantum action.
In Figure~\ref{fig4} a squashed bubble with $(A,B)=(2,0.5)$ is deformed with
the same $b$. The bubble is unstable against this deformation.

\vskip1mm
\vskip2mm
\noindent
\textbf{Acknowledgement:}
We thank H. Dorn for critical remarks.
This work was supported by the Swiss National Science Foundation.

\par\vfill
\def \tr { \varDelta ^{(d)}  }
\def \xai {x _{|i }^{\al }}
\def \xbj {x _{|j }^{\beta }}
\def \R {{\cal R} ^{(d)}}
\def \Rab {{\cal R}^{(d)}_{ab}}
\def \Raa {{\cal R}^{(d)}_{aa}}
\def \Rrrr {{\cal R}^{(d)}_{rr,r}}
\def \Ri {{\cal R} ^{(d-1)}}
\def \Riab {{\cal R} ^{(d-1)}_{ab} }
\def \Riaa {{\cal R} ^{(d-1)}_{aa} }
\def \Ric {{\cal R} ^{(d)}_{\mu \nu}}
\def \Riem {{\cal R}^{(d)}_ {\sigma \mu \rho \nu} }

\begin{appendix}
\section{Conventions}\label{sec:appendixa}
Let the $d$-dimensional manifold $\M $ with boundary $\RM $
be embedded a in space-time $X$. The metric on $\M$ is $g_{\al\beta}$.
The definition of the curvature terms on $\M $ is
\begin{equation}
{\mathcal R}^{\al } _{\beta \gamma \delta } =\;
\Gamma^{\al} _{\delta \beta , \gamma }-\Gamma ^{\al} _{\gamma\beta,\delta}
+\Gamma^{\si} _{\delta \beta }\Gamma ^{\al} _{\gamma \si }
-\Gamma ^{\si} _{\gamma \beta }\Gamma ^{\al} _{\delta \si }\label{A.1}
\end{equation}
and
\begin{equation}
{\mathcal R}_{\beta \delta }={\mathcal R}^{\al}_{\beta \al \delta}=
\Gamma^{\al}_{\delta\beta,\al}-\Gamma ^{\al} _{\al \beta , \delta }
+\Gamma ^{\si} _{\delta \beta }\Gamma ^{\al} _{\al \si }
-\Gamma ^{\si} _{\al \beta }
\Gamma ^{\al} _{\delta \si }.\label{A.2}
\end{equation}
The geometric properties of the boundary are usually described in terms
of induced curvatures. Choosing local coordinates $x^\alpha,\;
\alpha=0,\dots,d-1,$ in $X$ the boundary can (locally) be
parametrized through functions
\begin{equation}
x^{\al }  = f^{\al }  (u^i ),\quad i=1,\dots,d-1, \label{A.3}
\end{equation}
and then the induced metric is defined by
\begin{equation}
\tilde g_{ij} : = g_{\al \beta} \; \xai \xbj \; . \label{A.4}
\end{equation}
Here $\xai $ is the covariant derivative defined by the
induced metric, but since the $x^{\al }$ are invariants
under coordinate transformations of the $u^i$,
it is just the ordinary derivative with respect to $u^i$.
$\xai $ is a tangent vector and hence
\begin{equation}
g_{\al \beta} \xai n^{\beta } =0 \; .\label{A.5}
\end{equation}
$n ^{\beta} $ is the unit inner pointing normal of $\RM $.
The second fundamental form of the surface is the symmetric tensor
\begin{equation}
K_{ij} \, := g_{\al \beta } n^{\beta } x_{|ij}^{\al } \;,
\label{A.6}
\end{equation}
which is equivalent to
\begin{equation}
K_{ij}= - \xai \xbj n_{\al ;\beta}
\; . \label{A.7}
\end{equation}
Here it is understood that the normal field $n$ is extended to
a neighbourhood of $\partial \M$ and that the covariant derivative
is then computed with the connection on $X$. The value of $n_{\al;\beta}$
on the boundary does not depend on the extension.
Let us introduce the extrinsic version $\chi _{\mu \nu }$ of the
second fundamental form
\begin{equation}
\chi ^{\al \beta} \; : = \xai \xbj K^{ij} \; . \label{A.8}
\end{equation}
Introducing the projector
\begin{equation}
h_{\al \beta }= \; g_{\al \beta} - n_{\al} n_{\beta} = g_{\al\gamma}
g_{\beta \delta} \; x^{\gamma }_{,i}
x^{\delta}_{,j}\; \tilde g ^{ij}  \label{A.9}
\end{equation}
it can be cast into the form
\begin{equation}
\chi _{\mu \nu}= - h^{\si }_{\mu}
h^{\ro }_{\nu} n_{\si ;\ro} \; . \label{A.10}
\end{equation}
and this form is convenient since it only involves the metric on $\M$,
the normal field $n$ and its covariant derivative.
It is easily shown that
\begin{equation}
\hbox{Tr} \chi =g^{\mu\nu}\chi_{\mu\nu}=\tilde g^{ij}K_{ij}=\hbox{Tr}K\;.
\label{A.11}
\end{equation}
Using the relation
\begin{equation}
\nabla ^{(d-1)}_ {\mu } =h_{\mu }^{\nu }\nabla ^{(d)}_ {\nu } \label{A.12}
\end{equation}
between the extrinsic covariant derivative and the covariant derivative
on $\M$ (the instrinsic one on $\partial\M$ is just the projection
of $\nabla^{d-1}$ on $\partial\M$) one can prove the Gauss equation
\begin{equation}
{\mathcal R }^{(d-1)}_{\al \beta \gamma \delta}=\;
h^{\al \si } h^{\beta \rho}
h^{\gamma \mu } h^{\delta \nu}
{\mathcal R }^{(d)}_{\si \rho \mu \nu}  + \chi _{\al \gamma}
\chi _{\beta \delta }
- \chi _{\al \delta} \chi _{\beta \gamma } \; . \label{A.13}
\end{equation}

\def \bzphi {b_2[\varphi,g_{\mu\nu}] }
\def \bdhaphi {b_{{3\over 2}}[\varphi,g_{\mu\nu}] }
\def \bephi {b_1[\varphi,g_{\mu\nu}] }
\def \bhaphi {b_{{1 \over 2}}[\varphi,g_{\mu\nu}] }
\def \azphi {a_2[\varphi,g_{\mu\nu}] }
\def \aephi {a_1[\varphi,g_{\mu\nu}] }

\section{Seeley-de Witt coefficients}
\label{sec:appendixb}
These are the volume Seeley-deWitt coefficients for massless
scalars which are relevant in $d\leq 4$-spacetime dimensions:
\begin{align}
\aephi =& {1\over (4\pi)^{d/2}}
\big(\ft{1}{6}-\xi \big) \R  \varphi \label{B.1}\\
\azphi =& {1\over (4\pi)^{d/2}}
{1\over 180} \Big(\Riem {\mathcal R}^{(d) \sigma  \mu \rho \nu }
-\Ric {\mathcal R }^{(d)\mu \nu}   \label{B.2}\\
&\hskip18mm +(6-30\xi )\tr \R +90(\ft{1}{6}-\xi )^2\, {\R}^2 \Big) \varphi
\; .\nonumber 
\end{align} 
For Dirichlet boundary conditions
$\Phi |_{\RM } =0$
the surface coefficients read
\begin{align}
\bhaphi &= {1\over (4\pi)^{d/2 }}
\Big(-{\sqrt \pi \over 2} \Big)\varphi \label{B.3}\\
\bephi &= {1\over (4\pi)^{d/2}}
\Big(\ft{1}{3}\trc \varphi -\ft{1}{2} \pnmu \Big)\label{B.4}\\
\bdhaphi &= {1\over (4\pi)^{d/2 }}
{\sqrt \pi \over 192} \Big(\big[ -3{(\trc )}^2 + 6\trcc
-4{\mathcal R}^{(d-1)}
+12(8\xi -1) \R \big] \varphi\nonumber\\ &
\hskip26mm +30 \trc \pnmu
-24 \nmu \nnu \nabla _{\nu } \nabla _{\mu } \varphi \Big) 
\label{B.5}\\
\bzphi &={1 \over (4\pi)^{d/2}}  {1\over 1890} 
\Big\{\big[ 80 \trccc -66 \trc \trcc +10 ({\trc })^3  \nonumber\\
&\hskip32mm+105 \R \trc - 21 \Ric \cmunu  - \ft{189}{2} \nmu \nabla _{\mu } \R 
\nonumber\\ 
&\hskip32mm -21 \Ric \nmu \nnu \trc +84 \Riem \nsi \nro \cmunu   \nonumber\\
&\hskip32mm + 630 \xi \nmu \nabla _{\mu } \R-630 \xi \trc \R \big] \varphi 
\nonumber\\ 
&\hskip28mm+45 \trcc \pnmu + 126 \trc \tri -9 (\trc )^2 \pnmu  &\nonumber\\
&\hskip28mm-\ft{315}{2} \R \pnmu -\ft{315}{2} n^{\mu } \nabla _{\mu } \tri 
\nonumber\\
&\hskip28mm+ 945 \xi \R \pnmu  \Big\}\; . \label{B.6}
\end{align}
\end{appendix}


\begin{thebibliography}{10}
\bibitem{1} G. Plunien, B. M\"uller and W. Greiner, {\sl Phys. Rep.}
{\bf 134} (1986) 87.
\bibitem{2} C.M. Bender and P. Hays, {\sl Phys. Rev.}
{\bf D14} (1976) 2622.
\bibitem{3} S. Blau, M. Visser and A. Wipf, {\sl Nucl. Phys.}
{\bf B310} (1988) 163.
\bibitem{4} J.L. Cardy and I. Peschel, {\sl Nucl. Phys.}
{\bf B300} (1988) 377.
\bibitem{5} O. Alvarez, {\sl Nucl. Phys.} {\bf B216} (1983) 125;\\
B. Durhuus, P. Olesen and J.L. Petersen, {\sl Nucl. Phys.}
 {\bf B198} (1982) 157.
\bibitem{6} R. Seeley, {\sl Amer. J. Math.} {\bf 91} (1969) 889;
\bibitem{7} J. Melmed, {\sl J. Phys.} {\bf A21} (1989) L113;
\bibitem{8} I.G. Moss, {\sl Class. Quant. Grav.} {\bf 6} (1989) 659;\\
J.S. Dowker and J.P. Schofield, {\sl J. Math. Phys.}
{\bf 31} (1990) 808;\\
I.G. Moss and J.S. Dowker, {\sl Phys. Lett.} {\bf B229} (1989) 261.
\bibitem{9} T.P. Branson and P.B. Gilkey, {\sl Commun. Part. Diff. Eq.}
{\bf 15} (1990) 2.
\bibitem{10} A. Dettki, {\sl Thesis LMU-Munich}, July 1990.
\bibitem{11} M. Reed and B. Simon, {\sl Methods of Modern Math. Phys.},
Vol.II, Academic Press (1975).
\bibitem{12} P. Gilkey, {\sl Invariance Theory, the Heat Equation
and the Atiyah Singer Index Theorem}, Publish or Perish (1984).
\bibitem{13} I.G. Avramidi, {\sl Phys. Lett.} {\bf B238} (1990) 92.
\bibitem{14} J.L Cardy, in {\sl Phase trans. and crit. phen.},
Vol.11, ed. C. Domb and J.L Lebowitz, Academic Press (1987).
\bibitem{15} A. Hurwitz and R. Courant, {\sl Allgemeine
Funktionentheorie und elliptische Funktionen}, Springer, (1929).
\bibitem{16} I. Affleck, {\sl Phys. Rev. Lett.} {\bf 56} (1986) 746.
H. Bl\"ote, J. Cardy and M. Nightingale, {\sl Phys. Rev. Lett.}
{\bf 56} (1986) 742.
\bibitem{17} G. Grubb, {\sl Functional Calculus of Pseudodifferential
Boundary Problems}, Birk-hauser (1986).
\bibitem{18} J. Eskin, {\sl Israel J. Math.} {\bf 22} (1975) 214.
\bibitem{19} H. Widom, {\sl Bull. Sc. Math.} {\bf 104} (1980) 19.
\bibitem{20} H. Widom, {\sl Asymptotic Expansion for Pseudodifferential
Operators on bounded domains}, Lecture Note 1152, Springer (1985).
\bibitem{21} A. Erd\'elyi et.al., {\sl Bateman manuscript project},
Vol.1, McGraw-Hill (1954).
\bibitem{22} T. Eguchi, P.B. Gilkey and A. Hanson, {\sl Phys. Rep.}
{\bf 66} (1980) 213.
\end{thebibliography}
\end{document}